\documentclass[twocolumn]{aastex63}

\usepackage{graphicx}
\usepackage[varg]{txfonts}
\usepackage{natbib}
\usepackage{color, soul}

\received{}
\revised{}
\accepted{}

\submitjournal{ApJ}

\shorttitle{Planetary transits at radio}
\shortauthors{Selhorst et al.}

\begin{document}

%\title{Planetary transits at radio wavelengths: the influence of the hot Jupiters atmosphere}

\title{Planetary transits at radio wavelengths: secondary eclipses of hot Jupiter extended atmospheres}

\correspondingauthor{Caius L. Selhorst}
\email{caiuslucius@gmail.com}

\author{Caius L. Selhorst}
\affiliation{NAT -  N\'ucleo de Astrof\'isica, Universidade Cruzeiro do Sul/Universidade Cidade de S\~ao Paulo, S\~ao Paulo, SP, Brazil}

\author{Cassio L. Barbosa}
\affiliation{Dept. de F\'{\i}sica, Centro Universitário FEI, S\~ao Bernardo do Campo, SP, Brazil}

\author{Paulo J. A. Sim\~oes}
\affiliation{CRAAM, Universidade Presbiteriana Mackenzie, S\~ao Paulo, SP 01302-907, Brazil}
\affiliation{SUPA School of Physics and Astronomy, University of Glasgow, G12 8QQ, UK}

\author{Aline A. Vidotto}
\affiliation{School of Physics, Trinity College Dublin, The University of Dublin, Dublin 2, Ireland}

\author{Adriana Valio}
\affiliation{CRAAM, Universidade Presbiteriana Mackenzie, S\~ao Paulo, SP 01302-907, Brazil}

\begin{abstract}

When a planet transits in front of its host star, a fraction of its light is blocked, decreasing the observed flux from the star. The same is expected to occur when observing the stellar radio flux. However, at radio wavelengths, the planet also radiates, depending on its temperature, and thus modifies the transit depths. We explore this scenario simulating the radio lightcurves of transits of hot-Jupiters, Kepler-17b and WASP-12b, around solar-like stars. We calculated the bremsstrahlung radio emission at 17, 100, and 400~GHz originated from the star, considering a solar atmospheric model. The planetary radio emission was calculated modelling the planets in two scenarios: as a blackbody or with a dense and hot extended atmosphere. In both cases the planet radiates and contributes to the total radio flux. For a blackbody planet, the transit depth is in the order of 2--4\% and it is independent of the radio frequency. Hot-Jupiters planets with atmospheres appear bigger and brighter in radio, thus having a larger contribution to the total flux of the system. Therefore, the transit depths are larger than in the case of blackbody planets, reaching up to 8\% at 17~GHz. Also the transit depth is frequency-dependent. Moreover, the transit caused by the planet passing behind the star is deeper than when the planet transits in front of the star, being as large as 18\% at 400~GHz. In all cases, the contribution of the planetary radio emission to the observed flux is evident when the planet transits behind the star.

\end{abstract}


\section{Introduction} 

The discovery of exoplanets dates back to 1992, when \cite{Wolszczan1992} reported the detection of 2 planets around pulsar PSR B1257+12, by means of precise timing measurements of radio pulses with the Arecibo radio telescope. Currently, the vast majority of the the over 4000 exoplanets discovered so far were made by the $Kepler$ space telescope mission \citep{Borucki10} using the transit method.

The development of the instrumentation to observe planetary transits allowed the improvement of the observations from a tool merely used for detection of new exoplanets, to a powerful technique for characterizing exoplanets. In the last two decades, it became possible to derive the mass, radius, temperature, and even to probe the physical and chemical components of the atmosphere of a transiting exoplanet \citep[see][for example]{Fulton2017,Sanchez2019}.

Most of the planetary transits have been observed in the UV-optical-infrared window. \cite{Poppenhaeger2013} reported the observation of a planetary transit of HD~189733b, a hot Jupiter orbiting a K0V star, in soft X-rays. So far, there are only reports of upper limits for possible transits observed at MHz radio wavelengths \citep{Smith:2009,Lecavelier:2013}. \cite{Selhorst2013} showed the feasibility of detection by simulating observations at 17 GHz of planets, with sizes varying from super Earth to hot Jupiter, crossing the stellar disk. Observations at radio wavelengths were restricted to the interaction between the parent star and the atmosphere of the exoplanet \citep[see for example][and  references therein]{Pope2019}. As  \cite{Selhorst2013} showed, the observation of planetary transits at radio wavelengths is a very promising tool, specially for magnetically active stars such as red dwarfs.

In this paper, we explore the high frequency radio regime, which is accessible with the JVLA (Karl G. Jansky Very Large Array) and  ALMA (Atacama Large Millimeter/submillimeter Array). We note that exoplanets have also been predicted to produce auroral emission in the low-frequency radio regime \citep{Farrell:1999, Griebmeier:2007, Vidotto:2010, Vidotto:2019}. The exoplanetary auroral emission, generated by electron cyclotron maser instabilities (ECMI), is believed to arise from the interaction between the magnetised stellar wind and the planet magnetic field, which would enhance the detectability of exoplanets. Such an emission would make the planet brighter than the stellar emission itself at lower frequencies  \citep[e.g.][]{Farrell:1999,Vidotto2015}.

This emission is believed to be similar to auroral emission observed in brown dwarfs \citep{Kao:2016} and M dwarfs \citep{Hallinan:2019, Llama:2018}. In the exoplanetary case, however, the frequency of emission is expected to be much lower, due to lower magnetic field strengths. For this reason, radio arrays such as LOFAR and, in the future Square Kilometre Array \citep[SKA; ][]{Dewdney2009},  are better suited to observe exoplanetary auroral emission \citep{Zarka:2015}.

There have been numerous unsuccessful attempts to observe exoplanetary radio emission at lower frequencies \citep[e.g.][]{Lazio:2007, Smith:2009, Hallinan:2013, Sirothia:2014} A possible, but yet unlikely, reason for this is that hot Jupiters are unmagnetised and thus they do not generate ECMI. Alternatively, the non-detections are due to  observations  not being sufficiently sensitive, or that they are conducted at frequencies that do not match the (cyclotron) frequency of emission \citep{Bastian:2000}. Other possibilities are that a dense planetary atmosphere \citep{Weber:2017} or a dense stellar wind \citep{Vidotto:2017, Kavanagh:2019} could suppress ECMI planetary emission. The stellar wind itself can absorb planetary radio emission, causing radio planets to be eclipsed by the wind of their host stars \citep{Kavanagh:2020}.
While, like us, these authors  computed  radio eclipses of exoplanets, \citet{Kavanagh:2020} did not discuss eclipses due to thermal emission from the planet. Here, on the other hand, we focus on radio transits and eclipses that are caused by thermal emission of the planet. The advantage of our study is that, contrary to ECMI studies, we do not need to prescribe an unknown planetary magnetic field strength to predict the frequency of the emission. Given the large frequencies we investigate here, planetary magnetic fields would need to be unreasonably high ($>6$~kG) to cause ECMI at such high frequencies. 

Recently, \citet{Vedantham:2020} reported LOFAR detection of low-frequency radio emission from the quite M dwarf GJ 1151, believed to be induced by a close-in terrestrial mass planet \citep{Pope:2020}. Investigation of radio emission of exoplanets, from low to high radio frequencies, is thus very timely. The effects of the interaction of the stellar emission with the dense and hot atmosphere of close planets at radio wavelengths are explored in this work.

Primary and secondary transit simulations are performed at three distinct radio frequencies, 17, 100, and 400~GHz, generated by bremsstrahlung emission.

We report the radio light curves produced by our simulations for such a planet transit in the next section, whereas the discussion and conclusions are presented in the following sections.

\section{Simulations}

The light curve at radio wavelengths of a solar-like star is calculated assuming that the radio emission  is similar to that observed for the Sun \citep{Liseau2016}. In this case,  the radio emission of the host star was estimated using the solar atmospheric model proposed by \cite*{Selhorst2005a} (hereafter referred to as the SSC model). The SSC model was chosen due to the very good agreement with the solar brightness temperature observed at various radio frequencies (1 -- 405~GHz), as well as its  good agreement with the limb brightening observed in  solar maps obtained at 17, 100, and 230~GHz \citep{Selhorst2005a,Selhorst2005b,Selhorst2019}.  

The transit simulations were made based on a model developed in \cite{Silva2003} where the 2D images of the star (with limb brightening in this case) and that of the planet (opaque or not) are created. The position of the planet in its orbit is calculated at a given time interval and the sum of all the pixels in the image of the star-planet systems composes the light curve.

While in \cite{Selhorst2013}, the transiting planet was taken as an opaque disc, in this work, two scenarios are considered to estimate the hot Jupiter radio contribution. In the first one, the orbiting planet is assumed to behave as a blackbody (BB) without an atmosphere, which means that the planet's brightness and effective temperatures are the same ($\rm T_{Bp}=T_{eff}$). Following the results obtained for planetary transits observed at X-rays \citep{Poppenhaeger2013}, the second scenario considers a planetary dense and hot atmosphere, which increases the planet size at radio wavelengths. Both approaches are described in more detail bellow.

\subsection{Stellar maps}

As mentioned above, the parent star was considered to be solar-like, thus the brightness temperature  was calculated using the SSC atmospheric model.  For simplicity, bremsstrahlung was considered to be the unique emission mechanism, where we neglected the gyro-resonance contribution and the refraction effects \citep[see ][]{Tan2015} to the quiet Sun brightness temperature ($\rm T_{BS}$) at radio frequencies greater than $\sim$5~GHz.
 
\begin{figure}[!ht]
\centering
\includegraphics[width=9cm]{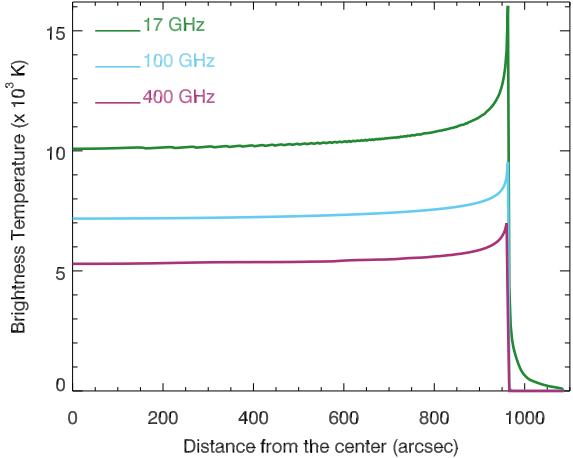} 
\caption{Center-to-limb variation of the stellar brightness temperature at 17, 100, and 400~GHz, plotted in green, blue, and magenta, respectively.}
\label{fig:limb}
\end{figure}
 
The brightness temperature of the stellar disc was calculated from spatially resolved solar observations at three radio frequencies: 17, 100, and 400~GHz. The emission at each frequency is formed at different atmospheric layers, with the higher frequencies originating closer to the photosphere. The brightness temperature for each frequency presents distinct values at the disc center and their images also display different limb brightening intensities.
 
 The center-to-limb variation of the brightness temperature at 17, 100, and 400~GHz is shown in Figure~\ref{fig:limb}  (green, blue, and magenta curves, respectively). While the 17~GHz emission originates from the upper chromosphere and presents a brightness temperature  raging  from $10\times 10^3$~K at disc center to a maximum of $16\times 10^3$~K at the limb, the emission observed at 100 and 400~GHz is produced in the atmospheric region close to the temperature minimum region, i.e., the region between the photosphere and chromosphere,  resulting in smaller temperatures, from $7.2\times 10^3$~K to $9.5\times 10^3$~K at 100~GHz and from  $5.3\times 10^3$~K to $7\times 10^3$~K at 400~GHz. 
 These center-to-limb profiles were used to generate bi-dimensional stellar maps. Each map was made with the same angular size of the Sun at 1 AU, with 3\arcsec\ pixel resolution (see Figure~\ref{fig:eclipse}a). This spatial resolution smoothed the limb brightening profiles. As a result, the position of the maximum value of the brightness temperature is the same at 17 and 100~GHz, 963\arcsec, whereas the limb maximum brightness temperature at 400~GHz occurs at 960\arcsec.            

\subsection{Blackbody planet}

While in the radio simulations performed by \cite{Selhorst2013} the planets eclipsing the star were considered as opaque discs, in our first approach in this work, the planets were simulated as blackbody objects without an extended atmosphere. Therefore, the planet brightness and effective temperatures are equal ($\rm T_{Bp}=T_{eff}$). We simulated the transit of two hot Jupiter planets, Kepler-17b and WASP-12b. Even though both planets orbit G-type stars, they present distinct sizes and temperatures. The planets parameters such as size, temperature, orbital period, and semi-major axis were obtained from the Extrasolar Planet Encyclopedia (\href{http://www.exoplanet.eu}{www.exoplanet.eu}) and are listed on Table~\ref{table0}.

\begin{table}
\caption{Characteristics of the planets used in the simulations}
\label{table0}
\begin{tabular}{ccccc}
\hline
Planet & Radius &  $\rm T_{eff}$ & Orbital Period  & Semi-Major \\
 & ($\rm R_S$) &  (K)  & (days) &  Axis (AU) \\
 \hline
Kepler-17b & 0.13 & 1655 & 1.4857 & 0.0259 \\
WASP-12b & 0.19 & 2593 & 1.0914 & 0.0234  \\
\hline
\end{tabular} 
\end{table}  

Both planets orbit very close to their host star, limiting the prospects of spatially resolving the system by future radio observations. Thus, the observed flux would be the total flux of the stellar system  $ F_{total}=F_S + F_p$, where $F_S$ and $F_p$  are, respectively, the star and the planet radio fluxes. During the transit of the planet in front of the star, a reduction of the total flux is expected due to the partial blockage of the stellar flux by the planet. In this case, the observed flux ($F$) is calculated as
\[F=\frac{F_{total}-F_{abs}}{F_{total}},
\]
where, $F_{abs}$ is the stellar radio flux blocked by planet.

\begin{figure*}[!ht]
\centering
\includegraphics[width=15cm]{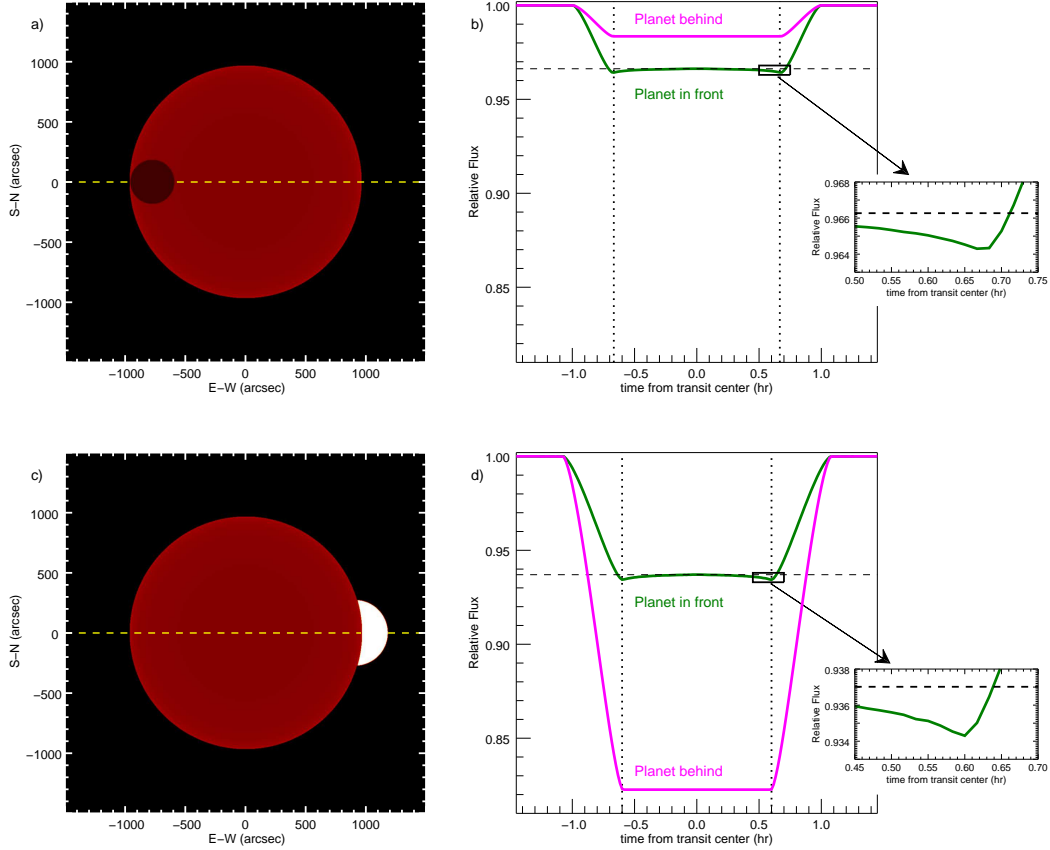} 
\caption{a) Stellar map simulated at 400~GHz, eclipsed by a planet with WASP-12b characteristics. The dashed line represents the path of the planet. (b) 400~GHz light curves obtained caused by the transit of the planet in front (green) and behind (magenta) the star. The dashed line is the flux reduction at the center of the transit, while the limb reduction bellow this value is showed in details in the right side box. The vertical  dotted lines were placed at the deepest transit positions.} c) Stellar map simulated at 400~GHz eclipsing a planet with WASP-12b characteristics and with a hot and dense atmosphere. (d) 400~GHz light curves obtained by the transit of the planet in front (green) and behind (magenta) the star.
\label{fig:eclipse}
\end{figure*}

An example of the stellar disc and the transiting planet is shown in Figure~\ref{fig:eclipse}a for WASP-12b at 400~GHz, where the dashed line represents the path of the planet. The resulting radio light curves are shown in Figure~\ref{fig:eclipse}b for two cases: the transit of the planet in front (green curve) and behind the star (magenta curve). Since the planets were simulated with uniform temperatures, when the planet hides behind the star, the light curves have flat intensities, as can be seen by the magenta curve in Figure~\ref{fig:eclipse}b. This behavior contrasts with the intensity variation in the light curve during the planetary transit in front of the star (green curve), that present a larger reduction at the edges caused by the planet blocking the limb brightening of the star, that are showed in detail in the right side boxes.

The flux reduction caused by the blockage of the limb brightening should be observed by the first and last points of the transit in which the planet is completely in front of the stellar disc (Figure~\ref{fig:eclipse}a). These transit positions are showed by dotted vertical lines in Figures~\ref{fig:eclipse}b and \ref{fig:eclipse}d. Due to the stellar limb brightening, the strongest flux reduction during the transit occurs when the planet is near the limb of the star. Note that for optical transits, the effect is opposite: due to limb darkening the strongest flux reduction happens during the center of the transit. In the cases of transits with large impact parameters, the transit duration would be shorter and thus its detection would require observations with short cadence.

Transit simulations were performed for the two planets, where the radio flux reduction of the transit lightcurve in each case was estimated at the frequencies of 17, 100, and 400~GHz. These flux reductions during transits reflect the planet contribution to the star-planet system flux.
For each case, the results are listed in the last 6 columns of Table~\ref{table2}. 
The transit simulation of the two planets showed a decrease in flux of less than 1.7\% for Kepler-17b, and less than 3.6\% for WASP-12b, which has a larger radius than Kepler-17b. As expected, the larger and hotter planet, WASP-12b, caused a larger flux reduction than that generated by Kepler-17b. 
 In the case of the transit light curve of Kepler-17b obtained considering the planet blackbody temperature, the decrease in flux of about 2\% is almost equal to that obtained with the earlier simulation done by \cite{Selhorst2013} considering opaque planets. 

The simulations also showed that the primary transit depth does not vary with frequency (see Table~\ref{table2}). On the other hand, the simulation of the secondary eclipse, when there is no contribution from the planet to the total flux, showed that the relative transit depth reduces with frequency, from 17 to 400~GHz, implying that the relative contribution of the planet radio emission to the total radio flux increases with frequency.

%\begin{table*}
%\caption{Flux Reduction Due to Transits of Blackbody Planets}
%\centering
%\label{table1}
%\begin{tabular}{ccccccc}
%\hline
%Planet  & \multicolumn{6}{c}{Largedt Flux Reduction}\\
%\hline
% & \multicolumn{3}{c}{Planet in front the Star}& \multicolumn{3}{c}{Planet behind the Star} \\
% \hline
%  & 17~GHz & 100~GHz & 400~GHz& 17~GHz & 100~GHz & 400~GHz \\
%\hline
%Kepler-17b &  1.7\% & 1.7\% & 1.7\% & 0.3\% & 0.4\% & 0.5\%\\
%WASP-12b & 3.6\% & 3.6\% & 3.6\% & 0.8\% & 1.2\% & 1.7\%\\
% Kepler-17b & 0.13 & 1655.0 & 1.4857 & 0.0259 & 1.73\% & 1.74\% & 1.74\% & 0.26\% & 0.37\% & 0.50\%\\
% WASP-12b & 0.19 & 2593.0 & 1.0914 & 0.0234 & 3.55\% & 3.56\% & 3.55\% & 0.84\% & 1.22\% & 1.65\%\\
%\hline
%\end{tabular} 
%\end{table*}  

\subsection{Hot Jupiter atmosphere}

To simulate the influence of hot Jupiter atmospheres in transit observations at radio frequencies, we considered a spherically symmetric atmospheric model, similar to the one proposed by \cite{Poppenhaeger2013}. Here, the atmosphere was considered fully ionized at the region where the radio emission is formed, with an electron density of $\rm n_e=7\times 10^9~cm^{-3}$ at $\rm 1.0~R_J$ above the visible disc of the planet, that is 10 times smaller than the used by \cite{Pope2019} in their estimations. The density scale height was kept equal to that one used in \cite{Pope2019}, that is $\rm H=5,000~km$. For simplicity, the temperature of the atmosphere was assumed to be constant, with a value of $\rm T=15,000~K$, at the region in which the radio emission is formed.

In our adopted atmospheric model the planetary atmosphere becomes optically thick at the selected radio frequencies. The bremsstrahlung opacity, $\kappa_\nu$, for a fully ionized plasma is proportional to $n_e^2\nu^{-3}$. As a consequence, the optical depth ($\tau_\nu=\int \kappa_\nu ds$) reaches values greater than unity at distinct atmospheric altitudes for the simulated radio frequencies. Although this makes the planet to appear larger at lower radio frequencies, as shown in Table~\ref{table2}, the difference between the smallest (400~GHz) and the biggest (17~GHz) radio planet is only $\rm0.03~R_S$, where $\rm R_S$ is the stellar radius, for both Kepler-17b and WASP-12b. Therefore, our modelled radio emission from the planets corresponds to a 15,000~K homogeneous brightness temperature. 

Once the model for the radio emission from the hot-Jupiter planets was developed, we calculated the transit lightcurves. Figure~\ref{fig:eclipse}c shows the transit of the hot WASP-12 planet and the star at 400~GHz. In this scenario, except for the sharpest edge of the limb brightening at 17~GHz (see Figure~\ref{fig:limb}), the brightness temperature of the planet is larger than the brightness temperature of the star. Thus the contribution of the planet to the total radio flux is larger than that in the blackbody case. This is due to the inclusion of the atmosphere, resulting in a higher brightness temperature and larger size of the planet. This effect can be seen in the larger transit depth obtained in the simulations for both Kepler-17b and WASP-12b planets. The lightcurves of the transit of the hot-Jupiter in front of the star (green) and behind it (magenta) are depicted in Figure~\ref{fig:eclipse}d for WASP-12b at 400~GHz.

The transit depth obtained from the simulations are listed on the bottom two rows of Table~\ref{table2} for transits of the planets in front of and behind the star.
Due to the similar planet sizes in radio, the planet's contribution to the total flux $F$ rises with frequency, since $F \propto \nu^2$. The flux reduction due the planetary transits presented distinct behaviours if the planet is in front or behind the star. 

When the planet transits in front of the star, the transit depth is larger at lower frequencies (Table~\ref{table2} columns 6--8), which is caused by the larger contribution of the stellar emission to the total flux (Figure~\ref{fig:size}). For Kepler-17b, the transit depth varies from 5.6 to 4.4\% from 17 to 400~GHz, whereas this variation is larger for WASP-12b, the depth decreases from 8 to 6.6\% from 17 to 400~GHz.

On the other hand, when the planet passes behind the star, the transit becomes deeper with increasing frequency (Table~\ref{table2}), reaching a flux reduction of 18\% at 400~GHz for WASP-12b. Moreover, because the brightness temperature of the planet is greater than that of the star, in all simulations with a hot planetary atmosphere, the transit is deeper when the planet passes behind the star (see Figure~\ref{fig:eclipse}d). 

For all frequencies simulated here, the hot Jupiters secondary transits were deeper than the primary ones. However, the differences between them reduces with decreasing frequency, while at 400~GHz the transit depth of WASP-12b varies from a primary transit with 6.6\% maximum reduction to a secondary with 17.7\% reduction, at 17~GHz the depth changed from 8.1\% to 11.5\%, for the primary and secondary transits, respectively. Moreover, assuming the models used in the simulations, the primary transit should became deeper than the secondary one when the stellar brightness temperature becames equal to the hot Jupiter one, which is obtained around 5~GHz in the SSC model \citep[see Figure 3 in ][]{Selhorst2005a}. Furthermore, for lower frequencies  typically formed at coronal heights, the primary transits tend to be much deeper than the secondary ones. 

\begin{table*}[t]
\caption{Flux Reduction Due to Transits of Blackbody and Hot-Jupiter Planets}
\label{table2}
\begin{tabular}{ccccccccccc}
\hline
Planet & $\rm T_{Bp}$ & \multicolumn{3}{c}{Radius in radio} & \multicolumn{6}{c}{Largest Flux Reduction}\\

 &   (K)  & \multicolumn{3}{c}{($\rm R_S$)} & \multicolumn{3}{c}{Planet in front of the Star}& \multicolumn{3}{c}{Planet behind the Star} \\
 \hline
 &  &  17~GHz & 100~GHz & 400~GHz & 17~GHz & 100~GHz & 400~GHz& 17~GHz & 100~GHz & 400~GHz \\
\hline
\multicolumn{11}{l}{Blackbody planets}\\
\hline
Kepler-17b & 1655 & 0.13 & 0.13 & 0.13 & 1.7\% & 1.7\% & 1.7\% & 0.3\% & 0.4\% & 0.5\%\\
WASP-12b & 2593 & 0.19 & 0.19 & 0.19  & 3.6\% & 3.6\% & 3.6\% & 0.8\% & 1.2\% & 1.6\%\\
\hline
\multicolumn{11}{l}{Hot-Jupiter atmosphere}\\
\hline
Kepler-17b & 15000 & 0.25 & 0.23 & 0.22 & 5.6\% & 4.9\% & 4.5\% & 7.8\% & 9.8\% & 11.9\%\\
WASP-12b & 15000 & 0.31 & 0.30 & 0.28 & 8.1\% & 7.4\% & 6.6\% & 11.5\% & 14.8\% & 17.7\%\\
% Kepler-17b & 15000.0 & 0.25 & 0.23 & 0.22 & 5.60\% & 4.92\% & 4.44\% & 7.81\% & 9.77\% & 11.90\%\\
% WASP-12b & 15000.0 & 0.31 & 0.30 & 0.28 & 8.09\% & 7.35\% & 6.57\% & 11.49\% & 14.81\% & 17.74\%\\
\hline
\end{tabular} 
\end{table*}  

\section{Discussion}

\begin{figure}[!ht]
\centering
\includegraphics[width=9cm]{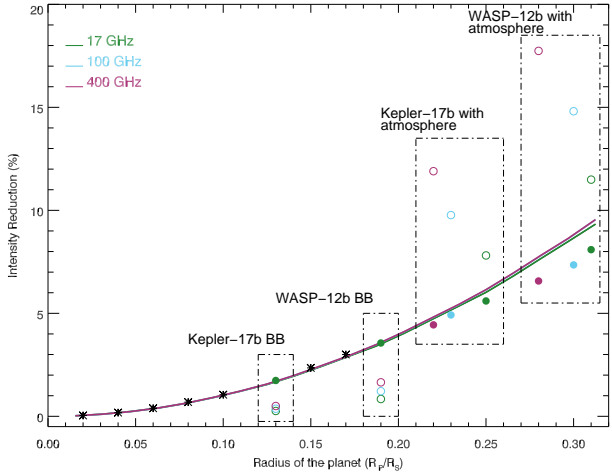} 
\caption{Transit depth of planets with different sizes at three distinct frequencies 17, 100, and 400~GHz, shown in green, blue and magenta, respectively. Following \cite{Selhorst2013}, the solid lines were obtained with the transit of opaque planets. The asterisks represent the results obtained by \cite{Selhorst2013} for different size planets. The results obtained for the blackbody planet model, and planets with an extended hot atmosphere (Table~\ref{table2}) are shown as circles, in which the filled circles represent the primary transit, whereas the open circles depict the secondary eclipse.}
\label{fig:size}
\end{figure}

To verify the consistency of the synthetic maps, the quiet star simulations performed by \cite{Selhorst2013} using the 17~GHz maps observed by the Nobeyama Radioheliograph were reproduced here with the synthetic maps instead of the observed ones. In these simulations, the planet was considered as an opaque disc. Figure~\ref{fig:size} shows the relative flux reduction during transits with planet size for three frequencies (17, 100, and 400 GHz).
The solid lines represent the simulations performed with the synthetic stellar maps based on the SSC model and the asterisks are the results obtained by \cite{Selhorst2013} using an observed quiet Sun map. In these simulations, the differences between results obtained at 17~GHz (green) and higher frequencies are very small. Moreover, the results obtained at 100~GHz (blue) and 400~GHz (magenta) lines are almost the same and cannot be distinguished in the plot.      

For comparison, the results obtained in this work are also plotted in Figure~\ref{fig:size} for Kepler-17b and WASP-12b. Both models of the blackbody (BB) planet and that with a hot extended atmosphere are shown in the rectangular boxes. The transit depths during primary transits (filled circles) and secondary eclipses (open circles) for the three radio frequencies 17 (green), 100 (blue), and 400~GHz (magenta) are plotted.

Clearly seen is the larger depth for the secondary eclipse compared to that of the primary transit, and how it increases for larger radio frequencies. As expected, the light curve intensity reduction is also larger for the hot extended planetary atmosphere because of the bigger size and flux contribution due to larger temperature.

In this paper, we suggest that the puffy atmospheres of close-in giant planets can be revealed in secondary transits in high-frequency radio observations.  Close-in exoplanets receive a large flux of energetic radiation from their host stars. In particular, the EUV and X-ray photons can photoionize hydrogen-dominated atmospheres, which are heated to temperatures that can reach above 10,000~K. As a consequence, these atmospheres expand and more easily evaporate. Planetary evaporation has been observed in several close-in planets in Ly-$\alpha$ transits \citep[e.g.][]{Vidal:2003, Kulow:2014, Bourrier:2016} and, more recently, are being revealed in the near-infrared He I triplet at 10833\AA\ \citep{Nortmann:2018, Spake:2018, Allart:2019}. Evaporation rates are higher in close-in planets that orbit active stars  \citep{Johnstone:2015, Kubyshkina:2018, Oklop:2019,  Allan:2019} due to their higher EUV and X-ray luminosities. Indirect evidence of atmospheric evaporation in close-in planets has also been seen in the distribution of planetary radii  \citep{Beauge:2013, Mazeh:2016, Fulton:2017}.

Here, we focus on the case of eclipsing planets that have spherically symmetric, dense and hot atmospheres. We note however that there is increased evidence both from observations \citep{Fossati:2010, Ehrenreich:2015} as well as from models \citep{Villareal:2014, Villarreal:2018, Matsakos:2015} that this is not always the case. Interactions with stellar winds and orbital motion predict that atmospheric material should be distributed asymmetrically around the planet, causing accretion streams, bow shocks and comet-like tails \citep[e.g.][]{Lai:2010, Vidotto:2010b, Bourrier:2013}. We leave the effects that asymmetric atmospheres have on radio lightcurves to a future study.

Our calculations assumed that planetary atmospheres do not vary in relatively short timescales. However, temporal variations in stellar activity are expected to lead to transit variability in timescales as short as a few transits apart up to stellar cycle timescales \citep{Vidotto:2011, Llama:2015}. Indeed, \citet{Lecavelier:2012} reported variations in the Ly-$\alpha$ transits of HD189733b separated by 1.5 yr. The transit variabilities were interpreted as being caused either by changes in stellar wind conditions or by the observed stellar flare, which could have increased high-energy irradiation and thus caused stronger evaporation. One way to overcome the impact of stellar variability in the interpretation of transit signals is by combining data from multiple transits \citep{Llama:2016}.

We investigated the possibility of detection of radio transits using current state-of-the-art and future radio telescopes, for the conditions simulated in this work. For 17 GHz, we used the curves of the sensitivity of future radio telescopes \citep[Figure 1 of ][]{Pope2019} as a guide. Considering a solar-type star, like $\epsilon$ Eri at its true distance (3.2 pc) and a flux reduction of 10\% during the transit, the transit would be barely detectable after 1 hour of integration time at JVLA. With the SKA1 telescope it would be possible to observe transits at a 5 minute cadence, which would allow the detection of the progression of the transit itself. At a distance of 20 pc, the 10\% transit for the same solar-type star would only be detected by the planned SKA2 extension, but only after more than 30 min of integration time.

For higher frequencies, we used the  sensitivity calculator to estimate the detection of a 10\% transit of such solar-type star. In this case, we used the same input parameters of \citet{Selhorst2013}: flux of 340 $\mu$Jy at 345 GHz with Bandwidth of 16 GHz and optimal observing conditions for 50 12-m antennas. To be able to detect the transit, the rms of the observation has to be within at least at 10\% of the flux, which means 34 $\mu$Jy. This could be achieved in less than 8 minutes of integration time, possibly allowing the detection of the shape of the transit curve.

\section{Conclusions}

In this work, we estimated the planet radio flux contribution relative to that of the star taking into consideration the planet atmosphere at three radio frequencies of 17, 100, and 400~GHz. Two models were analysed, a blackbody planet and a planet with an extended atmosphere with a brightness temperature of 15,000 K. 
The planetary flux contribution was estimated measuring the transit depth during the secondary eclipse.

Simulations of the light curve during the primary and secondary transits of hot Jupiters around solar type stars at radio frequencies were made.  The star was simulated with the solar atmospheric model SSC \citep{Selhorst2005a} at 17, 100, and 400~GHz. 
The physical characteristics of Kepler-17b and WASP-12b were used in the simulations, where the two atmospheric scenarios were considered. In the first one, the planets were considered as black bodies, which implies that the planet brightness and effective temperatures were equal, moreover, the radio planet had the same size as observed at optical wavelengths. 

Since future radio observations will detect the whole stellar system (star+planet), the blackbody planet contribution is low, causing a decrease of $<4\%$ in the total flux during transits. This result  is similar to that obtained by \cite{Selhorst2013} for  opaque planets. However, the consideration of the planet temperature (about 1650 K and 2600 K for Kepler-17b and WASP-12b, respectively) resulted in a secondary transit reflecting 
the planet larger contribution to the total radio flux (see Table~\ref{table2}).    

In the second approach, the effect of the planetary extended atmosphere was simulated using the characteristics proposed by \cite{Poppenhaeger2013} to reproduce the X-rays transit profiles. The hot and dense planetary atmosphere increases the size of Kepler-17b from 0.13~$\rm R_S$ to 0.22--0.25~$\rm R_S$, whereas, WASP-12b grew from  0.19~$\rm R_S$ to 0.28--0.31~$\rm R_S$. In both cases, the smallest and largest sizes where obtained at 400 and 17~GHz, respectively. Due to the radio planet increase, the primary transit became deeper, with a depth varying from 5.6 to 4.4\% from 17 to 400~GHz for Kepler-17b, whereas for WASP-12b, the depth decreased from 8 to 6.6\% from 17 to 400~GHz.

Since the atmosphere is dense enough to became optically thick and the plasma temperature was 15,000~K, the planet brightness temperature is greater than that expected for the star at the simulated frequencies. Nevertheless, due to the larger star size, the stellar flux is still greater than the planetary one. However, the radio planet contribution to the system total flux increased considerably reaching approximately 8--12~\% for Kepler-17b and 11--18~\% for WASP-12b,  increasing in frequency from 17 to 400~GHz. The radio planet contribution could be obtained by the deep flux reduction during the planet secondary transit, i.e., when the planet is totally eclipsed by the star.

While \cite{Pope2019} studied the possibility of a planetary transit to be observed at low radio frequencies by the SKA, here we presented an analysis at higher frequencies (17, 100, and 400~GHz), which could be currently observed by the JVLA and ALMA. In comparison with \cite{Selhorst2013}, the inclusion of a planetary atmosphere increased the possibility of primary planetary transit be observed at radio frequencies due to the increase of the planet radius. Moreover, our results showed new observational possibilities to verify the planetary flux through secondary transits, which is a promising tool in the millimeter and submillimeter range.

\acknowledgments

The authors wish to thank the anonymous referee for her/his comments and suggestions that improved the revised version of the paper. C.L.S., P.J.A.S., and A.V. acknowledge financial support from the S\~ao Paulo Research Foundation (FAPESP), grant numbers 2019/03301-8 and 2013/10559-5.  A.A.V. acknowledges funding from the European Research Council (ERC) under the European Union's Horizon 2020 research and innovation programme (grant agreement No 817540, ASTROFLOW).

\bibliographystyle{aasjournal}
\bibliography{references} 

\end{document}